\documentclass[nofootinbib,twocolumn, prd, preprintnumbers]{revtex4-1}
\usepackage{amsmath,amssymb,graphicx,bm,psfrag,color,slashed,subcaption,array,subfiles,cancel}
\usepackage[version=4]{mhchem}
\usepackage[colorlinks=true, allcolors=blue]{hyperref}
\usepackage[dvipsnames]{xcolor}

\usepackage{siunitx} 
\DeclareSIUnit{\year}{y}

\begin{document}

\preprint{IPPP/23/35}
\vspace{0.5cm}
\title{Fifth forces from QCD axions scale differently}
\author{Martin Bauer}
\author{Guillaume Rostagni}

\address{Institute for Particle Physics Phenomenology, Department of Physics\\
Durham University, Durham, DH1 3LE, United Kingdom}

\begin{abstract}
We reexamine the low-energy potential for a macroscopic fifth force generated from the exchange of two axions. The shift-symmetry of the linear axion interactions leads to a potential falling off as $V(r)\sim 1/r^5$. We find that in the case of the QCD axion higher-order terms in the Lagrangian break the shift symmetry and lead to the dominant contribution to the potential scaling as $V(r)\sim 1/r^3$.  These terms are generated by the same physics responsible for the axion mass and therefore the new contributions to the potential induce a different force for external nucleons and leptons. We demonstrate how this result affects the sensitivity of searches for new long-range forces.

\end{abstract}

\maketitle

\section{Introduction}
Fifth forces have been identified as potential probes for axions very early~\cite{Moody:1984ba}. The early focus was on the exchange on spin-dependent interactions, which are the consequence of the exchange of single light CP-odd scalars. The leading contribution to a spin-independent long-range force mediated by axions is generated by axion-pair exchange at one loop~\cite{Grifols:1994zz} and therefore similarly suppressed as the 'neutrino-force' generated by the exchange of neutrino pairs~\cite{Feinberg:1968zz, Hsu:1992tg}. The potential corresponding to the exchange of 
a pair of neutrinos scales as $V(r)\sim 1/r^5$, as does the potential generated by the exchange of pairs of massless axions~\cite{Grifols:1994zz,Ferrer:1998ju}.
In contrast, the exchange of pairs of pseudoscalars leads to a non-relativistic potential scaling as $V(r)\sim 1/r^3$, whereas the potential for an axion-Higgs portal scales as $V(r)\sim 1/r^7$~\cite{Ferrer:1998rw,Bauer:2022rwf}. The difference between the potentials induced by pseudoscalars and axions is a consequence of the manifest shift-symmetry that protects all linear axion interactions and have been discussed already in the case of the pion in very early literature~\cite{Dyson:1948a,Lepore:1952a,Drell:1952a,Drell:1953a}. \\
In this letter we show that in the case of the QCD-axion the dominant contribution to the potential is generated by the same physics responsible for the axion mass and that these contributions generate a $V(r)\sim 1/r^3$ potential even though they are induced by higher-order operators in the effective field theory (EFT) expansion in the axion decay constant. Since the axion mass is generated by strong dynamics these additional contributions to the low-energy potential only occur for external hadrons. Axion-induced forces between leptons as well as between hadrons and leptons are substantially weaker. This could allow to directly measure the contribution to the axion mass from the chiral anomaly by comparing different searches for fifth forces. \\

There are several ways to search for the effects of a new, macroscopic force including searches with Cavendish-type experiments~\cite{Adelberger:2006dh}, searches for new forces in atoms and molecules~\cite{Safronova:2017xyt}, measurements of the effective Casimir pressure~\cite{Chiu:2010ybt,Bezerra:2014dja} and experiments specifically designed to suppress the Casimir force~\cite{Chen:2014oda}. We introduce the different contributions to axion interactions at low energy in Section~\ref{sec:axioninteractions}, derive the potential for axions including the new contributions from high-order operators in Section~\ref{sec:axionforce}, and demonstrate the effect of the new contribution for the Casimir-less experiment~\cite{Chen:2014oda} in Section~\ref{sec:constraints}.

\section{Two axion interactions}
\label{sec:axioninteractions}
The Lagrangian for an axion interacting with fermions can be written in a form that is explicitly shift invariant apart from the axion mass $m_a$,
\begin{align}\label{eq:Lagdell}
\mathcal{L}=\frac{1}{2}(\partial a)^2-\frac{m_a^2}{2}a^2-\sum_\psi\frac{c_\psi}{2}\frac{\partial_\mu a}{f}\bar \psi \gamma_5 \gamma^\mu \psi\,, 
\end{align}
up to linear order in the axion field over the axion decay constant. This Lagrangian can be rewritten by using the divergence of the axial-vector current 
\begin{align}\label{eq:Lagm}
\mathcal{L}&=\frac{1}{2}(\partial a)^2-\frac{m_a^2}{2}a^2\notag\\
&\quad+\sum_\psi c_\psi i m_\psi\frac{a}{f}\bar \psi \gamma_5 \psi-c_\psi\frac{\alpha Q_\psi^2}{4\pi}\frac{a}{f}F_{\mu\nu}\tilde F^{\mu\nu},
\end{align}
where we assume the fermions only carry electric charge, otherwise there would be additional couplings to gauge bosons. Even though \eqref{eq:Lagdell} and \eqref{eq:Lagm} are both linear in $a/f$ they lead to contradicting results for processes with more than one axion involved. The reason is that the divergence of the axial-vector current or equivalently the equations of motion for the axion only capture terms up to linear order in the fields. A consistent rescaling of the fermion fields generates higher order terms in $a/f$ that precisely account for the difference between results obtained from  \eqref{eq:Lagdell} and \eqref{eq:Lagm} (details are given in Appendix~\ref{apprescaling}). The effects can be accounted for by modifying the anomaly equation for the divergence of the axial-vector
current
\begin{align}\label{eq:anomaly}
\frac{c_\psi}{2}\frac{\partial_\mu a}{f}\bar \psi \gamma_5 \gamma^\mu \psi &=-c_\psi i m_\psi\frac{a}{f}\bar \psi \gamma_5 \psi +c_\psi^2 m_\psi\frac{a^2}{f^2}\bar \psi \psi\notag\\
&\quad +c_\psi\frac{ \alpha Q_\psi^2}{4\pi}\frac{a}{f}F_{\mu\nu}\tilde F^{\mu\nu}+\mathcal{O}\Big(\frac{a^3}{f^3}\Big)\,.
\end{align}
To quadratic order in the axion fields the inclusion of the additional operator in \eqref{eq:Lagm} restores the results obtained using the shift invariant coupling. However the shift invariance in \eqref{eq:Lagdell} is explicitly broken by the presence of an axion mass. Treating $m_a^2$ as the only spurion that breaks the shift invariance suggests the existence of higher order shift symmetry breaking operators
\begin{align}\label{eq:massspurion}
\mathcal{L}_\text{ssb}\ni \sum_\psi c_{m}\frac{m_a^2 a^2}{f^3}\bar \psi \psi\,.
\end{align}
These operators spoil the cancellation in \eqref{eq:anomaly}.
In general it is a conservative assumption that the spurion is given by $m_a^2$, because the source of shift symmetry breaking responsible for generating the axion mass can induce higher-order operators that are less suppressed than \eqref{eq:massspurion}. An example of such an enhancement is the coupling of the QCD axion to nucleons. The shift symmetry is broken by the presence of light quark masses and the QCD confinement scale. Interactions between the QCD axion and nucleons are therefore shift-invariant or suppressed by these spurions. At leading order the operators of the two-flavor chiral Lagrangian coupling baryons to pions and axions are
\begin{align}\label{eq:chiral1}
    \mathcal{L}^{(1)}=\bar N \left( i\slashed D -m_N+\frac{g_A}{2}\gamma^\mu\gamma^5 u_\mu +g_0\gamma^\mu\gamma^5 a_\mu^{(s)}\right)N\,.
\end{align}
Couplings to the axion enter via the covariant derivative and the vielbeins $u_\mu$ and $a_\mu^{(s)}$, which both contain the axion in an explicitly shift-invariant way~\cite{Vonk:2020zfh, Bauer:2021mvw}. 

At second order there are four operators
\begin{align}\label{eq:chiral2}
\mathcal{L}^{(2)}&=c_1\text{tr}[\chi_+]\bar N N -\frac{c_2}{4m^2}\text{tr}[u_\mu u_\nu](\bar N D^\mu D^\nu N + \text{h.c.})\notag\\
&+\frac{c_3}{2}\text{tr}[u_\mu u^\mu]\bar N N -\frac{c_4}{4}\bar N \gamma^\mu\gamma^\nu [u_\mu,u_\nu]N\,.
\end{align}
All operators in $\mathcal{L}^{(2)}$ are shift-invariant apart from the operator with coefficient $c_1$, which contains a shift-symmetry breaking interaction 
\begin{align}\label{eq:chiraloperator}
c_1\text{tr}[\chi_+]\bar N N= c_{N}\frac{a^2}{f^2} \bar N N + \ldots
\end{align}
The axion field enters via 
\begin{align}
\!\!\chi_+&= 2B_0 \big(\xi^\dagger m_q(a) \xi^\dagger+\xi m_q^\dagger(a) \xi\big)\,,\\
\!\!m_q(a)&=e^{-i\kappa_q \frac{a}{2f}(2c_{GG}+c_u+c_d)}m_q e^{-i\kappa_q \frac{a}{2f}(2c_{GG}+c_u+c_d)}\!,
\end{align}
where $\xi=\exp(i/\sqrt2\,\Pi/f_\pi)$ contains the pion fields, the quark masses read $m_q=\text{diag}(m_u,m_d)$, $\kappa_q=\text{diag}(\kappa_u,\kappa_d)$ are unphysical parameters subject to the constraint $\kappa_u+\kappa_d=1$, and $c_{GG}$ denotes the axion coupling to gluons 
\begin{align}
\mathcal{L}\ni c_{GG}\frac{\alpha_s }{4\pi }\frac{a}{f}G_{\mu\nu}\tilde G^{\mu\nu}\,.
\end{align}
After rotating into the mass eigenbasis and taking into account contributions from pion mixing one can write the leading terms for the amplitude of axions coupled to nucleons from \eqref{eq:chiral1} and \eqref{eq:chiral2} as
\begin{align}
&\!\!\!i\mathcal{A}(N(k')\to N(k)+a(q))=-\frac{g_{N}}{4f}\bar u_N(k')\slashed{q}\gamma_5 u_N(k),\\
&\!\!\!i\mathcal{A}(N(k')\to N(k)+2a(q/2))=-\frac{c_N}{f^2}\bar u_N(k')u_N(k),
\end{align}
respectively. Here, the couplings are defined for protons and neutrons $N=p,n$, as
\begin{align}\label{eq:nucleoncouplings}
g_{p/n}&=g_0(c_u+c_d+2c_{GG})\notag\\
&\pm g_A \frac{1}{1-\tau_a^2}\left(c_u-c_d+2c_{GG} \frac{m_d-m_u}{m_u+m_d}\right)\,,\notag\\
c_N&=c_1 \frac{m_\pi^2}{2} \frac{4c_{GG}^2(1-\tau_a)^2 +(c_{u}-c_{d})^2\tau_a^2}{(1-\tau_a)^2}\,,
\end{align}
where $\tau_a=m_a^2/m_\pi^2$. The iso-scalar and iso-vector coupling constants are determined using lattice gauge theory~\cite{Liang:2018pis,FlavourLatticeAveragingGroup:2019iem} $g_0 =
0.440(44)$ and experimentally extracted from nucleon beta decay~\cite{ParticleDataGroup:2020ssz} $g_A = 1.2754(13)$, respectively.  The low energy coefficients $c_1, c_2, c_3, c_4$ can be found in~\cite{Alarcon:2012kn} and we use $c_1= -1.26(14)$ GeV$^{-1}$ here. The axion couplings to gluons and quarks in \eqref{eq:nucleoncouplings} are to be evaluated at the QCD scale~\cite{Chala:2020wvs, Bauer:2020jbp}.

Expanding $c_N$ in small axion masses and using the expression for the QCD axion mass with $m_u=m_d$ one can write the coefficient in \eqref{eq:massspurion} as $c_m=-8c_1 f^3/f_\pi^2$, which corresponds to a substantial enhancement compared with the naive assumption.

Since the axion has a potential, in principle any quadratic interaction can also give rise to a linear spin-independent interaction \emph{if} the axion vacuum expectation value doesn't vanish. The Vafa-Witten theorem guarantees that $\langle a\rangle =0$ in vacuum~\cite{Vafa:1984xg}, but in a high density environment the potential is modified and $\langle a\rangle=a_0\neq 0$, leading to long-range forces for large, dense objects such as neutron stars~\cite{Hook:2017psm, Balkin:2020dsr}. Linear interactions proportional to the theta angle are strongly suppressed~\cite{Chang:1985mu, Haber:1987nx, Mantry:2014eya}. For the remainder of this paper we focus on the spin-independent force induced by the exchange of axion pairs.
The importance of the shift-symmetry breaking operator has been pointed out previously in the context of coherent axion-nucleon scattering~\cite{Fukuda:2021drn}.

\begin{figure}[tbp]
    \centering
\includegraphics[width=.5\textwidth]{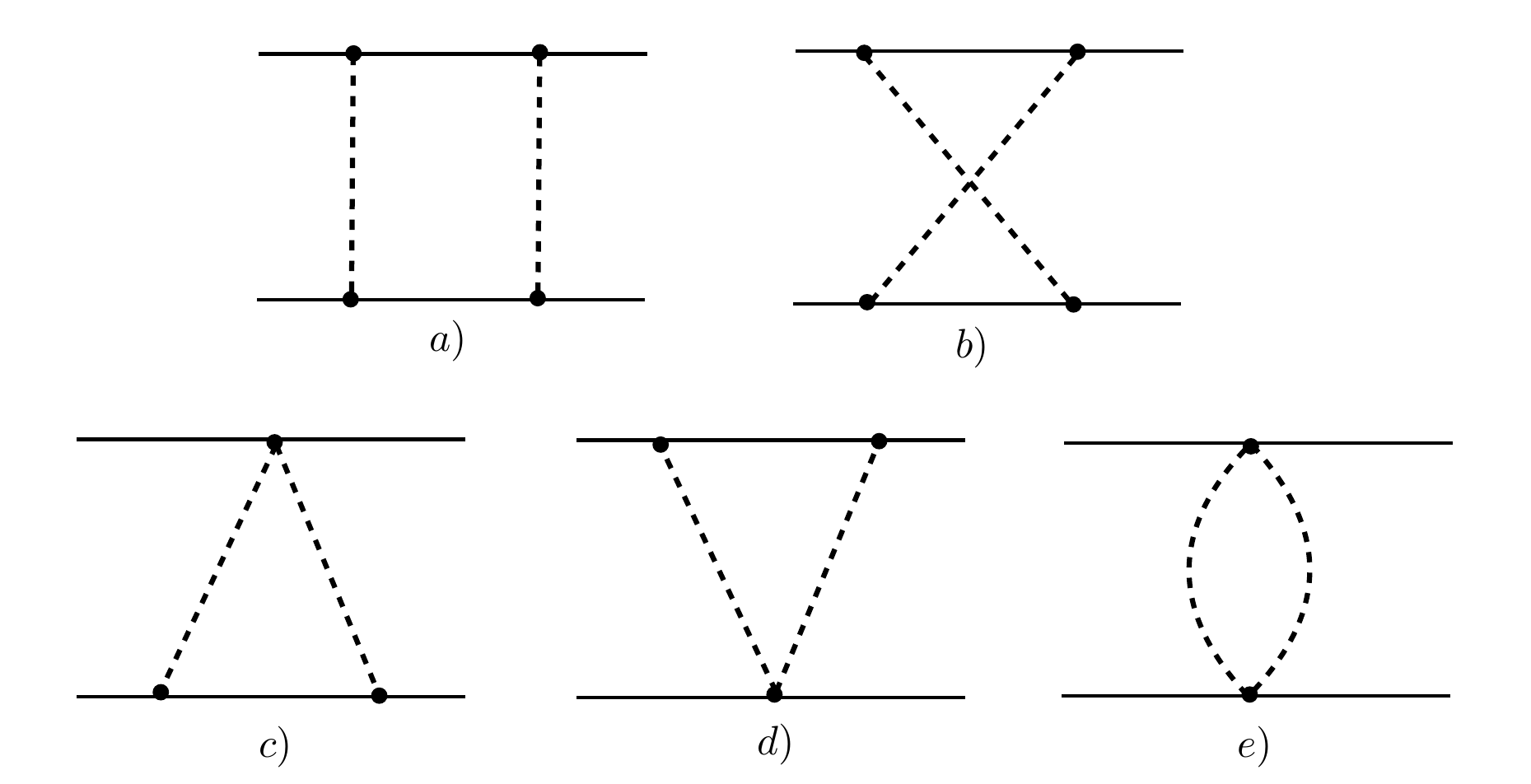}
    \caption{Diagrams contributing to the potential generated by two-axion exchange from linear interactions $a)$ and $b)$, from linear and quadratic interactions $c)$ and $d)$ and from purely quadratic interactions in $e)$.  }
    \label{fig:diagrams}
\end{figure}

\section{The axion force}
\label{sec:axionforce}
In the following we will derive the potential for the spin-independent force induced by the exchange of a pair of axions~\cite{Grifols:1994zz, Ferrer:1998rw, Ferrer:1998ue}. We show explicitly that the contributions from the linear and quadratic axion interactions in \eqref{eq:anomaly} cancel and that the shift-symmetry breaking interaction induced by \eqref{eq:chiraloperator} spoils this cancellation and provides the most important contribution to the potential.
We obtain the non-relativistic potential for the exchange of two axions can be obtained by taking the discontinuities in the scattering amplitude in the non-relativistic
limit and perform the Fourier transform. Feynman diagrams for the two-axion exchange are shown in Fig.~\ref{fig:diagrams}. In the basis with derivative axion-interactions \eqref{eq:Lagdell} only the diagrams $a)$ and $b)$ contribute. We instead use the non-derivative basis for which one needs to include diagrams $c)$, $d)$ and $e)$, taking into account the quadratic axion coupling in \eqref{eq:anomaly} to obtain a consistent result. Operators breaking the shift invariance generate additional contributions to $c), d)$ and $e)$.

In the heavy fermion limit and retaining only terms odd in the momentum exchanged $\sqrt t$ in the amplitudes\footnote{Terms even in $\sqrt t$ are cancelled by the contribution from the iterated single-axion exchange potential in the massless pseudoscalar limit \cite{Ferrer:1998ue}, we assume this is still the case for a massive pseudoscalar.} we obtain the following spin-independent contributions at next-to-leading order in $m_a$ for diagrams \ref{fig:diagrams} $a)$ and $b)$ 
\begin{align} \label{eq:potential_ab}
    V_{ab}&(r) = -\frac{c_{\psi_1}^2 c_{\psi_2}^2}{64\pi^3 f^4} m_{\psi_1} m_{\psi_2} \bigg\{ \frac{1}{r^3}  x_a K_1(x_a)  \nonumber \\
     &+  \bigg( \frac{1}{m_{\psi_1}^2} + \frac{1}{m_{\psi_2}^2} -\frac{1}{2m_{\psi_1} m_{\psi_2}}\bigg)\frac{3}{r^5}\notag\\
     &\quad\times\bigg[ \left( x_a + \frac{x_a^3}{6}  \right) K_1(x_a) + \frac{x_a^2}{2}  K_0(x_a) \bigg] \bigg\}
\end{align}

in which we define the dimensionless variable $x_a=2m_ar$ and $K_0(x_a)$ and $K_1(x_a)$ are modified Bessel functions of the second kind. In the case of a pseudoscalar particles described by the linear coupling in \eqref{eq:Lagm} the potential \eqref{eq:potential_ab} would be the full potential and one recovers the leading term
\begin{equation}\label{eq:Vabexp}
    V_{ab}(r) = -\frac{c_{\psi_1}^2 c_{\psi_2}^2}{64\pi^3 f^4}   \frac{m_{\psi_1} m_{\psi_2}}{r^3}  + O\left(\frac{m_a^2}{r^3},\frac{1}{r^{5}}\right) \, .
\end{equation}
The contributions from diagrams $c)$ and $d)$ are given by
\begin{align}
    V_c(r) &= \frac{c_{\psi_1}^2 c_{\psi_2}^2}{64\pi^3 f^4} m_{\psi_1} m_{\psi_2} \bigg\{ \frac{1}{r^3}  x_a K_1(x_a)  \nonumber \\
    & + \frac{1}{m_{\psi_2}^2} \frac{3}{r^5} \bigg[  \left( x_a + \frac{x_a^3}{6}  \right) K_1(x_a)  +  \frac{x_a^2}{2} K_0(x_a) \bigg] \bigg\}\,,\notag\\[2pt]
     V_d(r)&  =V_c(r)  \quad \text{with}\quad   m_{\psi_1}\leftrightarrow m_{\psi_2}
\end{align}
whereas diagram $e)$ gives
\begin{equation}
    V_e(r) = -\frac{c_{\psi_1}^2 c_{\psi_2}^2}{64 \pi^3 f^4} m_{\psi_1} m_{\psi_2} \frac{1}{r^3}   x_a K_1(x_a)  \, .
\end{equation}
In the sum of these contributions the terms proportional to $r^{-3}$ cancel out and we are left with
\begin{align}\label{eq:fullpotential}
 V(r) &= V_{ab}(r)+V_{c}(r)+V_d(r)+V_e(r)\\
 &=\frac{3c_{\psi_1}^2 c_{\psi_2}^2}{128\pi^3 f^4}\frac{1}{r^5}\bigg[\left( x_a + \frac{x_a^3}{6}  \right) K_1(x_a) + \frac{x_a^2}{2}  K_0(x_a) \bigg] \notag
     \end{align}
Expanding this result around $x_a = 0$ we recover the familiar $r^{-5}$ potential
\begin{equation}\label{eq:Vfullexp}
    V(r) = \frac{3c_{\psi_1}^2 c_{\psi_2}^2}{128\pi^3 f^4} \left[ \frac{1}{r^5} - \frac 13 \frac{m_a^2}{r^3} + O(m_a^4) \right] \,.
\end{equation}
In the case of axions with an explicit mass term the potential~ \eqref{eq:fullpotential} is proportional to $V(r)\sim 1/r^5$ up to terms suppressed by $m_a^2 < 1/r^2$ as a result of the shift symmetry of the Lagrangian. Additional contributions from shift-symmetry breaking operators~\eqref{eq:massspurion} are suppressed by $\sim 1/f^6$. However, in the case of the QCD axion there are additional terms at the same order in $1/f^4$ induced by the quadratic interaction terms~\eqref{eq:chiraloperator} proportional to the shift-symmetry breaking spurion responsible for the axion mass. Evaluated for a potential between two nucleons $N_1$ and $N_2$ the additional diagrams generate the potential
\begin{align} \label{eq:potential_spur}
    V_\text{sp.}(r) &= \frac{1}{64 \pi^3 f^4}   \bigg\{ -c_{N_1} c_{N_2}\frac{1}{r^3} x_a K_1(x_a)\notag\\
    &+ \frac{3}{4} \left[    c_{N_1} g_{N_2}^2\frac{1}{m_{N_2}} +   c_{N_2}g_{N_1}^2\frac{1}{m_{N_1}}  \right]  \frac{1}{r^5} \nonumber\\
    &\quad \times\bigg[\left( x_a + \frac{x_a^3}{6}  \right) K_1(x_a) + \frac{x_a^2}{2}  K_0(x_a) \bigg]\bigg\}\,\notag\\
    &=-\frac{c_{N_1} c_{N_2}}{64\pi^3 f^4}   \frac{1}{r^3}  + O\left(\frac{m_a^2}{r^3},\frac{1}{r^{5}}\right)\,, 
\end{align}
where $g_N$ and $c_N$ are defined in~\eqref{eq:nucleoncouplings}. The contributions from the quadratic axion interaction induced by the spurion  dominate over the contribution from the interaction induced by shift-invariant operators even though the latter appear at leading order in the EFT expansion. Note that this is different from the corrections in the expansion~\eqref{eq:Vfullexp} which are suppressed by the axion mass, which in the case of the QCD axion scales as $m_a^2 \propto f_\pi^4/f^2$. While \eqref{eq:fullpotential} results in a repulsive potential, \eqref{eq:potential_spur} can in principle have either sign, but is universally attractive for a QCD axion only interacting with gluons. The effect of the shift-symmetry breaking interaction -to leading order- doesn't affect leptons, because Feynman diagrams $c)$ and $d)$ in Fig.~\ref{fig:diagrams} don't contribute to the leading term in \eqref{eq:potential_spur}.

\begin{figure}
    \centering
    \includegraphics[scale=0.5]{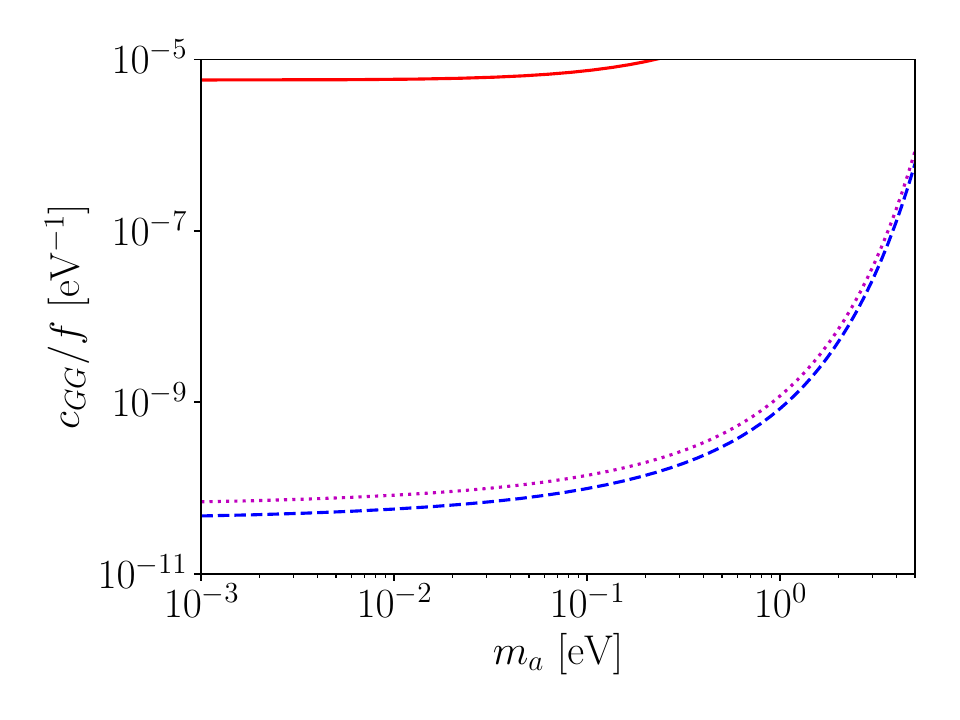}
    \caption{Limits on the axion-gluon couplings $c_{GG} /f$ obtained from the Casimir-less experiment \cite{Chen:2014oda}. The dashed blue contour corresponds to the limit obtained using the pseudoscalar form of the potential~\eqref{eq:Vabexp}, the solid red contour corresponds to the axion potential without shift-symmetry breaking terms ~\eqref{eq:fullpotential}, and the dotted purple line corresponds to the full axion potential including the nucleon spurion term~\eqref{eq:potential_spur}. }
    \label{fig:casless-plot}
\end{figure}

\section{Fifth force constraints on QCD Axions}
\label{sec:constraints}
In the following we demonstrate the effect of the shift-symmetry breaking interaction on the sensitivity of experiments searching for a fifth force. We consider the simplest QCD axion model with a single coupling to gluons described by the Wilson coefficient $c_{GG}$ keeping it's mass $m_a$ a free parameter. Bounds from atomic and molecular spectroscopy aren't substantially changed by the inclusion of the higher order operators~\eqref{eq:chiral2} because the leading effects only affect nucleon-nucleon interactions. We instead consider experiments probing macroscopic, spin-independent forces such as the one described in \cite{Chen:2014oda} in which the difference in the force between a sphere and a plate of two different materials is probed, which minimises the contribution from the Casimir effect. The accuracy in measuring this force (or absence thereof) has been used in \cite{Klimchitskaya:2015zpa} to obtain the best limits on the pseudoscalar-to-nucleon coupling in the meV--eV range for an experiment of this type.\\
The corresponding differential force between a sphere of radius $R$ and a disk with thickness $D$ with Au and Si coating placed at a distance $\ell$ from the sphere reads
\begin{align}\label{eq:Fa}
\Delta F (\ell)&= 2\pi C_s \left[C_\text{Au}-C_{\text{Si}}\right] \!\int_{\ell}^{2R+\ell}\!\!\!\!\!\!dz_1(R^2-(z_1-R-\ell)^2)\notag\\
&\times\!\frac{\partial}{\partial z_1}\!\int_{-D}^0\!\!dz_2\!\int_0^\infty \!\!\rho d\rho\, \mathcal{V}\big(\sqrt{\rho^2+(z_1-z_2)^2}\big)\,,
\end{align}
where we factor out the coupling constants such that $V(r)=c_{\psi_1}^2c_{\psi_2}^2 \mathcal{V}(r)$ in the case of the potential derived from the derivative interaction~\eqref{eq:fullpotential} and $V_\text{sp}(r)=c_{N_1}c_{N_2} \mathcal{V}_\text{sp}(r)$ for the leading term of \eqref{eq:potential_spur} and define the material-dependent prefactors in terms of the axion couplings to nucleons~\eqref{eq:nucleoncouplings}
\begin{align}\label{eq:CXgpgn}
C_X&=\rho_X\left(\frac{g_p^2}{4}\frac{Z_X}{m_X}+\frac{g_n^2}{4}\frac{N_X}{m_X} \right)\quad\text{for}\quad\eqref{eq:fullpotential}\,,\\
\label{eq:CXcncn}C_X&=\rho_X c_{N}\frac{A_X}{m_X} \quad\text{for}\quad\eqref{eq:potential_spur}\,,
\end{align}
with density $\rho_X$, average number of protons and neutrons $A_X=Z_X+N_X$ and mean masses $m_X$ of the disc and sphere atoms for a material $X$. The calculation of~\eqref{eq:Fa} is lengthy but straightforward following \cite{Bezerra:2014dja,Klimchitskaya:2015zpa,Klimchitskaya:2021lak} and we give the result in Appendix~\ref{apppotential}.
As expected the axion force derived from the shift-symmetry breaking interactions $V_\text{sp}(r)$ grows with $\ell$ compared to the force derived from derivative interactions as $\Delta F_\text{sp}/\Delta F(\ell)\sim m_\pi^2\ell^2$.

The resulting bounds on $c_{GG}/f$ are shown as a function of the axion mass in Fig.~\ref{fig:casless-plot}. The dashed blue line is the bound obtained by using the interaction Lagrangian in~\eqref{eq:Lagm}. However, including the quadratic interaction terms as shown in~\eqref{eq:anomaly} changes the potential to the $1/r^5$ form in~\eqref{eq:fullpotential}, resulting in a substantially weaker bound compared to the previous potential as shown by the position of the solid red line in the figure.
The quadratic nucleon spurion term in~\eqref{eq:chiraloperator} generates the additional potential~\eqref{eq:potential_spur} proportional to $1/r^3$ at leading order. Furthermore, the coupling of this term is not suppressed by the axion mass unlike what one would expect from a spurion breaking the axion shift symmetry. As a result this spurion term generates a bound close to that obtained using the $1/r^3$ pseudoscalar potential; this bound is shown by the dotted purple line in Fig.~\ref{fig:casless-plot}. 
Importantly, this contribution to the low-energy potential is only relevant if the external states are nucleons. As a result similar probes for fifth forces are only sensitive to the leading term in the potential if they don't depend on the axion coupling to leptons. e.g. atomic or molecular spectroscopy~\cite{Karshenboim:2010ck, Jaeckel:2010xx, Salumbides:2013dua,Salumbides:2013aga,Salumbides:2015qwa, Jones:2019qny} is only sensitive to the $1/r^5$ contribution to the potential, whereas measurements of cold neutron scattering or bouncing Neutrons are sensitive to the dominant $1/r^3$ part of the potential~\cite{Frank:2003ms,Nesvizhevsky:2007by, Kamiya:2015eva, Brax:2011hb, Brax:2013cfa}. We'll provide a more comprehensive discussion in a companion paper. 


\section{Conclusions}
\label{sec:conclusions}
We identify the dominant contribution to the low-energy potential for the macroscopic fifth force incuded by axion pair exchange for axions that -like the QCD axion- interact with gluons and thus obtain part of their mass from the chiral anomaly. This contribution arises from higher-order operators of the axion Lagrangian that would be naively expected to produce subleading effects. We show explicitly that these operators not only generate the most important contribution to the low-energy potential but result in a scaling of the non-relativistic potential $V(r) \sim 1/r^3$ as opposed to the leading term $V(r) \sim 1/r^5$ expected from derivative interactions. Moreover, since to the QCD axion mass is generated via strong dynamics, this new contribution is only present for interactions between nucleons and so the nature of the shift-symmetry breaking for an axion can be probed via the comparison of different searches for fifth forces. We demonstrate the impact at the example of a Casimir-less fifth-force experiment and find an improved sensitivity of almost 5 orders of magnitude.

\appendix
\section{}\label{apprescaling}\vspace{-1mm}

The axion interaction with chiral fermions $\psi=\psi_L+\psi_R$ in the UV theory can be derived from the Lagrangian 
\begin{align}
\mathcal{L}_\text{UV}=\frac{1}{2}\bar\psi i\overset{\leftrightarrow}{\slashed{\partial}}\psi-y \bar\psi_L S \psi_R +h.c
\end{align}
after the scalar$S$ developes a vacuum expectation value $f$ such that the fermion mass is given by $m=yf$ and 
\begin{align}
S=(f+s)\exp\Big(2i\frac{a}{f}\Big)
\end{align}
with a scalar field $s$ and the Goldstone boson $a$. Ignoring interactions of the scalar mode the Lagrangian reads
\begin{align}
\mathcal{L}=\frac{1}{2}\bar\psi i\slashed{\partial}\psi-m \bar\psi_L \exp\Big(2i\frac{a}{f}\Big)  \psi_R +h.c
\end{align}
For small $a/f$ one can expand the exponent and obtains interactions
\begin{align}
\mathcal{L}=\frac{1}{2}\bar\psi i\overset{\leftrightarrow}{\slashed{\partial}}\psi-m\left(2i \frac{a}{f}-2\frac{a^2}{f^2}+\mathcal{O}\left(\frac{a^3}{f^3}\right)\right) \bar\psi_L \psi_R +h.c
\end{align}
Alternatively one can rescale the fermion fields 
\begin{align}
\psi_L\to \exp\left(i\frac{a}{f}\right)\psi_L,\quad \psi_R\to \exp\left(-i\frac{a}{f}\right)\psi_R\,,
\end{align}
and find instead the explicitly shift invariant Lagrangian
\begin{align}\label{eq:explscale}
\mathcal{L}=\frac{1}{2}\bar\psi i\overset{\leftrightarrow}{\slashed{\partial}}\psi - \frac{\partial_\mu a}{f}\bar\psi\gamma_5\gamma^\mu \psi-m\bar\psi_L \psi_R +h.c
\end{align}
This leads to a non-relativistic potential scaling like $1/r^5$. Often this interaction term is rewritten using the equation of motion of the fermion fields 
\begin{align}\label{eq:eomlag}
\mathcal{L}&=\frac{1}{2}\bar\psi i\overset{\leftrightarrow}{\slashed{\partial}}\psi - \frac{a}{f}\Big(\bar\psi\gamma_5\overset{\rightarrow}{\slashed{\partial}} \psi+\bar\psi\gamma_5\overset{\leftarrow}{\slashed{\partial}} \psi\Big)-m\bar\psi_L \psi_R +h.c\notag\\
&=\frac{1}{2}\bar\psi i\overset{\leftrightarrow}{\slashed{\partial}}\psi +2im \frac{a}{f}\bar\psi\gamma_5\psi-m\bar\psi_L \psi_R +h.c
\end{align}
This form of the Lagrangian leads to different Feynman rules and for example the non-relativistic potential from two axion exchange has a $1/r^3$ dependence, because higher-order terms aren't captured by the naive application of the equations of motion. Instead, we rescale the fermion fields with field dependent factors that are linear in the axion field $L,R\propto a$ and factors that are quadratic in the axion fields $N,S\propto a^2$ such that
\begin{align}
\psi_L+L \psi_L + N \psi_L\,,\\
\psi_R+R \psi_R + S \psi_R\,.
\end{align}
We then find to linear order in $a$:
\begin{align}\label{eq:LR}
\mathcal{L}(a)&=\frac{1}{2}\big(L+L^\dagger\big)\bar\psi_L i\overset{\leftrightarrow}{\slashed{\partial}}\psi_L 
+\frac{1}{2}(\partial_\mu L -\partial_\mu L^\dagger)\bar \psi_L i\gamma^\mu \psi_L\notag\\
&+\frac{1}{2}\big(R+R^\dagger\big)\bar\psi_R i\overset{\leftrightarrow}{\slashed{\partial}}\psi_R+\frac{1}{2}(\partial_\mu R -\partial_\mu R^\dagger)\bar \psi_R i\gamma^\mu \psi_R \notag
\\
&- \frac{a}{f}\left(\bar\psi\gamma_5\overset{\rightarrow}{\slashed{\partial}} \psi+\bar\psi\gamma_5\overset{\leftarrow}{\slashed{\partial}} \psi\right)
\notag\\
&-m(L^\dagger+R)\bar \psi_L \psi_R -m(L+R^\dagger)\bar \psi_R \psi_L\,.
\end{align}
Applying the equations of motions for the axion is equivalent to the choice $L=ia/f$ and $R=-i a/f$. For this choice the first term vanishes and the terms in line 2 and 3 in \eqref{eq:LR} cancel and the remaining term reads
\begin{align}
\mathcal{L}(a)=2mi\frac{a}{f}\psi \gamma_5 \psi\,,
\end{align}
in agreement with \eqref{eq:eomlag}.
Now we consistently shift the terms quadratic in $a$ and find
\begin{align}
\mathcal{L}(a^2)&=\frac{1}{2}\big(LL^\dagger+N+N^\dagger\big)\bar\psi_L i\overset{\leftrightarrow}{\slashed{\partial}}\psi_L \notag\\
&+\frac{1}{2}(L^\dagger\partial_\mu L -L\partial_\mu L^\dagger+\partial_\mu N-\partial_\mu N^\dagger)\bar \psi_L i\gamma^\mu \psi_L\notag\\
&+\frac{1}{2}\big(RR^\dagger+S+S^\dagger\big)\bar\psi_R i\overset{\leftrightarrow}{\slashed{\partial}}\psi_R\notag\\
&+\frac{1}{2}(R^\dagger\partial_\mu R -R\partial_\mu R^\dagger+\partial_\mu S-\partial_\mu S^\dagger)\bar \psi_R i\gamma^\mu \psi_R \notag\\
&+ \frac{a}{f}\bigg[(L+L^\dagger)\Big(\bar\psi_L\overset{\rightarrow}{\slashed{\partial}} \psi_L+\bar\psi_L\overset{\leftarrow}{\slashed{\partial}} \psi_L\Big)
\notag\\
&+(\partial_\mu L + \partial_\mu L^\dagger) \bar \psi_L\gamma^\mu \psi_L \\
&-(R+R^\dagger)\Big(\bar\psi_R\overset{\rightarrow}{\slashed{\partial}} \psi_R+\bar\psi_R\overset{\leftarrow}{\slashed{\partial}} \psi_R\Big)
\notag\\
&-(\partial_\mu R + \partial_\mu R^\dagger) \bar \psi_R\gamma^\mu \psi_R\Big]\notag\\
&-m(L R^\dagger+N^\dagger+S)\bar \psi_R \psi_L -m(LR^\dagger+S^\dagger+N)\bar \psi_R \psi_L\,.\notag
\end{align}

Choosing $N$ and $S$ to be real eliminates every term apart from the first, third and last line. Setting 
\begin{align}
2N=2S=-L^\dagger L =-R^\dagger R=-\frac{1}{2}\frac{a^2}{f^2}
\end{align}
cancels the terms with derivative interactions and yields
\begin{align}
\mathcal{L}(a^2)=2m \frac{a^2}{f^2}\bar \psi \psi\,.
\end{align}
Including this operator in the calculation of the non-relativistic potential cancels terms that scale as $1/r^3$ and reproduces the $1/r^5$ potential obtained from the explicitly scale invariant form \eqref{eq:explscale}.
\vspace{-2mm}
\section{}\label{apppotential}\vspace{-1mm}

Integrating \eqref{eq:Fa} for the potential \eqref{eq:fullpotential} and \eqref{eq:potential_spur} yields respectively
\begin{align}
\Delta F(\ell) &= \frac{3 }{64\pi m_a} \frac{1}{f^4}  | C_\text{Au} - C_\text{Si}|  \\
&\times \int_1^\infty \mathrm du \frac{\sqrt{u^2 - 1}}{u^3} \sum_l C_l \Psi(m_a u) \,, \label{eq:cas_fA}\nonumber\\
\Delta F_\text{sp}(\ell) &= \frac{1}{32\pi m_a}   \frac{1}{f^4} | C_\text{Au} - C_\text{Si}| \int_1^\infty \mathrm du \\
&\times \frac{\sqrt{u^2 - 1}}{u^3} e^{-2m_a u \ell} \left( 1 - e^{-2m_a u D} \right) X(m_a u) \nonumber\label{eq:cas_fPS} 
\end{align}
where $\ell$ is the separation between the sphere and the surface of the disc with thickness $D$. The coefficients $C_\text{X}$ are given by \eqref{eq:CXgpgn} and \eqref{eq:CXcncn}, respectively, and the function $X(x)$ is given by eq (11) in \cite{Klimchitskaya:2015zpa}. The experiment measured the differential force between either Au or Si sectors of a rotating disc.

\newpage  The sphere is made of sapphire (sa.) coated with Au and Cr, and the sum in~\eqref{eq:cas_fA} is
\begin{align}
    \sum_l C_l \Psi(x) &= C_\text{Au} \Psi\left( x; R,r \right) \nonumber \\[-8pt]
    &+ \left( C_\text{Cr} - C_\text{Au} \right)  \Psi\left( x; R - d_\text{Au}, r + d_\text{Au} \right) \nonumber\\
    &+ \left( C_\text{sa.} - C_\text{Cr} \right) \nonumber \\
    &\times \Psi\left( x; R - d_\text{Au} - d_\text{Cr}, r + d_\text{Au} + d_\text{Cr} \right) 
\end{align}
with $R$ the radius of the sphere, $d_\text{Au}$ and $d_\text{Cr}$ the thicknesses of the gold and chrome coatings, and the function
\begin{align}
    \Psi(x;R_l,r_l) &= 
    8x^4 \int_{r_l}^{2R_l + r_l} \mathrm dz\  \left[ R_l^2 - (R_l + r_l - z)^2 \right] \nonumber \\
    &\times \Bigg\{ -\frac{e^{-2xz }}{2xz } \left( 1  - \frac{z }{D + z} e^{-2x D} \right) \nonumber\\
    &+ \text{Ei} \left[ -2x(D + z) \right] - \text{Ei} \left[ -2xz \right] \Bigg\}  \,,
\end{align}
where Ei$(x)$ is the exponential integral function.


\begin{thebibliography}{99}

\bibitem{Moody:1984ba}
J.~E.~Moody and F.~Wilczek,
Phys. Rev. D \textbf{30} (1984), 130


\bibitem{Ferrer:1998rw}
F.~Ferrer and M.~Nowakowski,
Phys. Rev. D \textbf{59} (1999), 075009
[arXiv:hep-ph/9810550 [hep-ph]].

\bibitem{Grifols:1994zz}
J.~A.~Grifols and S.~Tortosa,
Phys. Lett. B \textbf{328} (1994), 98-102
[arXiv:hep-ph/9404249 [hep-ph]].

\bibitem{Feinberg:1968zz}
G.~Feinberg and J.~Sucher,
Phys. Rev. \textbf{166} (1968), 1638-1644

\bibitem{Hsu:1992tg}
S.~D.~H.~Hsu and P.~Sikivie,
Phys. Rev. D \textbf{49} (1994), 4951-4953
[arXiv:hep-ph/9211301 [hep-ph]].

\bibitem{Ferrer:1998ju}
F.~Ferrer, J.~A.~Grifols and M.~Nowakowski,
Phys. Lett. B \textbf{446} (1999), 111-116
[arXiv:hep-ph/9806438 [hep-ph]].

\bibitem{Ferrer:1998rw}
F.~Ferrer and M.~Nowakowski,
Phys. Rev. D \textbf{59} (1999), 075009
[arXiv:hep-ph/9810550 [hep-ph]].

\bibitem{Bauer:2022rwf}
M.~Bauer, G.~Rostagni and J.~Spinner,
Phys. Rev. D \textbf{107} (2023) no.1, 015007
[arXiv:2207.05762 [hep-ph]].



\bibitem{Lepore:1952a}
J.~V.~Lepore,
Phys. Rev. \textbf{88} (1952) no.4, 750.

\bibitem{Drell:1952a}
S.~D.~Drell and E.~M.~Henley,
Phys. Rev. \textbf{88} (1952) no.5, 1053.

\bibitem{Drell:1953a}
S.~D.~Drell and K.~Huang,
Phys. Rev. \textbf{91} (1953) no.6, 1527.

\bibitem{Dyson:1948a}
F.~J.~Dyson,
Phys. Rev. \textbf{73} (1948) no.8, 929.

\bibitem{Adelberger:2006dh}
E.~G.~Adelberger, B.~R.~Heckel, S.~A.~Hoedl, C.~D.~Hoyle, D.~J.~Kapner and A.~Upadhye,
Phys. Rev. Lett. \textbf{98} (2007), 131104
[arXiv:hep-ph/0611223 [hep-ph]].

\bibitem{Safronova:2017xyt}
M.~S.~Safronova, D.~Budker, D.~DeMille, D.~F.~J.~Kimball, A.~Derevianko and C.~W.~Clark,
Rev. Mod. Phys. \textbf{90} (2018) no.2, 025008
[arXiv:1710.01833 [physics.atom-ph]].

\bibitem{Bezerra:2014dja}
V.~B.~Bezerra, G.~L.~Klimchitskaya, V.~M.~Mostepanenko and C.~Romero,
Eur. Phys. J. C \textbf{74} (2014), 2859
[arXiv:1402.3228 [hep-ph]].

\bibitem{Chiu:2010ybt}
H.~C.~Chiu, G.~L.~Klimchitskaya, V.~N.~Marachevsky, V.~M.~Mostepanenko and U.~Mohideen,
Phys. Rev. B \textbf{81} (2010) no.11, 115417
[arXiv:1002.3936 [quant-ph]].

\bibitem{Chen:2014oda}
Y.~J.~Chen, W.~K.~Tham, D.~E.~Krause, D.~Lopez, E.~Fischbach and R.~S.~Decca,
Phys. Rev. Lett. \textbf{116} (2016) no.22, 221102
[arXiv:1410.7267 [hep-ex]].

\bibitem{Vonk:2020zfh}
T.~Vonk, F.~K.~Guo and U.~G.~Mei\ss{}ner,
JHEP \textbf{03} (2020), 138
[arXiv:2001.05327 [hep-ph]].

\bibitem{Bauer:2021mvw}
M.~Bauer, M.~Neubert, S.~Renner, M.~Schnubel and A.~Thamm,
JHEP \textbf{09} (2022), 056
[arXiv:2110.10698 [hep-ph]].


\bibitem{Liang:2018pis}
J.~Liang, Y.~B.~Yang, T.~Draper, M.~Gong and K.~F.~Liu,
Phys. Rev. D \textbf{98} (2018) no.7, 074505
[arXiv:1806.08366 [hep-ph]].



\bibitem{FlavourLatticeAveragingGroup:2019iem}
S.~Aoki \textit{et al.} [Flavour Lattice Averaging Group],
Eur. Phys. J. C \textbf{80} (2020) no.2, 113
[arXiv:1902.08191 [hep-lat]].

\bibitem{ParticleDataGroup:2020ssz}
P.~A.~Zyla \textit{et al.} [Particle Data Group],
PTEP \textbf{2020} (2020) no.8, 083C01

\bibitem{Alarcon:2012kn}
J.~M.~Alarcon, J.~Martin Camalich and J.~A.~Oller,
Annals Phys. \textbf{336} (2013), 413-461
[arXiv:1210.4450 [hep-ph]].




\bibitem{Chala:2020wvs}
M.~Chala, G.~Guedes, M.~Ramos and J.~Santiago,
Eur. Phys. J. C \textbf{81} (2021) no.2, 181
[arXiv:2012.09017 [hep-ph]].

\bibitem{Bauer:2020jbp}
M.~Bauer, M.~Neubert, S.~Renner, M.~Schnubel and A.~Thamm,
JHEP \textbf{04} (2021), 063
[arXiv:2012.12272 [hep-ph]].

\bibitem{Vafa:1984xg}
C.~Vafa and E.~Witten,
Phys. Rev. Lett. \textbf{53} (1984), 535

\bibitem{Hook:2017psm}
A.~Hook and J.~Huang,
JHEP \textbf{06} (2018), 036
[arXiv:1708.08464 [hep-ph]].





\bibitem{Balkin:2020dsr}
R.~Balkin, J.~Serra, K.~Springmann and A.~Weiler,
JHEP \textbf{07} (2020), 221
[arXiv:2003.04903 [hep-ph]].


\bibitem{Chang:1985mu}
D.~Chang, R.~N.~Mohapatra and S.~Nussinov,
Phys. Rev. Lett. \textbf{55} (1985), 2835

\bibitem{Haber:1987nx}
H.~E.~Haber and M.~Sher,
Phys. Lett. B \textbf{196} (1987), 33-38

\bibitem{Mantry:2014eya}
S.~Mantry, M.~Pitschmann and M.~J.~Ramsey-Musolf,
[arXiv:1411.2162 [hep-ph]].

\bibitem{Fukuda:2021drn}
H.~Fukuda and S.~Shirai,
Phys. Rev. D \textbf{105} (2022) no.9, 095030
[arXiv:2112.13536 [hep-ph]].

\bibitem{Ferrer:1998ue}
F.~Ferrer and J.~A.~Grifols,
Phys. Rev. D \textbf{58}, 096006 (1998)
[arXiv:hep-ph/9805477 [hep-ph]].

\bibitem{Klimchitskaya:2015zpa}
G.~L.~Klimchitskaya and V.~M.~Mostepanenko,
Eur. Phys. J. C \textbf{75} (2015) no.4, 164
[arXiv:1503.04982 [hep-ph]].

\bibitem{Klimchitskaya:2021lak}
G.~L.~Klimchitskaya and V.~M.~Mostepanenko,
Universe \textbf{7} (2021) no.9, 343
[arXiv:2109.06534 [hep-ph]].




\bibitem{Salumbides:2013dua}
E.~J.~Salumbides, W.~Ubachs and V.~I.~Korobov,
J. Molec. Spectrosc. \textbf{300} (2014), 65
[arXiv:1308.1711 [hep-ph]].

\bibitem{Salumbides:2013aga}
E.~J.~Salumbides, J.~C.~J.~Koelemeij, J.~Komasa, K.~Pachucki, K.~S.~E.~Eikema and W.~Ubachs,
Phys. Rev. D \textbf{87} (2013) no.11, 112008
[arXiv:1304.6560 [physics.atom-ph]].

\bibitem{Salumbides:2015qwa}
E.~J.~Salumbides, A.~N.~Schellekens, B.~Gato-Rivera and W.~Ubachs,
New J. Phys. \textbf{17} (2015) no.3, 033015
[arXiv:1502.02838 [physics.atom-ph]].



\bibitem{Karshenboim:2010ck}
S.~G.~Karshenboim,
Phys. Rev. D \textbf{82} (2010), 073003
[arXiv:1005.4872 [hep-ph]].

\bibitem{Jaeckel:2010xx}
J.~Jaeckel and S.~Roy,
Phys. Rev. D \textbf{82} (2010), 125020
[arXiv:1008.3536 [hep-ph]].


\bibitem{Jones:2019qny}
M.~P.~A.~Jones, R.~M.~Potvliege and M.~Spannowsky,
Phys. Rev. Res. \textbf{2} (2020) no.1, 013244
[arXiv:1909.09194 [hep-ph]].




\bibitem{Nesvizhevsky:2007by}
V.~V.~Nesvizhevsky, G.~Pignol and K.~V.~Protasov,
Phys. Rev. D \textbf{77} (2008), 034020
[arXiv:0711.2298 [hep-ph]].

\bibitem{Frank:2003ms}
A.~Frank, P.~van Isacker and J.~Gomez-Camacho,
Phys. Lett. B \textbf{582} (2004), 15-20
[arXiv:nucl-th/0305029 [nucl-th]].

\bibitem{Kamiya:2015eva}
Y.~Kamiya, K.~Itagami, M.~Tani, G.~N.~Kim and S.~Komamiya,
Phys. Rev. Lett. \textbf{114} (2015), 161101
[arXiv:1504.02181 [hep-ex]].

\bibitem{Brax:2011hb}
P.~Brax and G.~Pignol,
Phys. Rev. Lett. \textbf{107} (2011), 111301
[arXiv:1105.3420 [hep-ph]].

\bibitem{Brax:2013cfa}
P.~Brax, G.~Pignol and D.~Roulier,
Phys. Rev. D \textbf{88} (2013), 083004
[arXiv:1306.6536 [quant-ph]].











\end{thebibliography}
\end{document}